**Electronic transport through carbon nanotubes -- effects of structural deformation and tube chirality**

Amitesh Maiti,[1,*] Alexei Svizhenko, [2,*] and M. P. Anantram[2]

[1]Accelrys Inc., 9685 Scranton Road, San Diego, CA 92121

[2]NASA Ames Research Center, Mail Stop: T27A-1, Moffett Field, CA 94035-1000

Abstract

Atomistic simulations using a combination of classical forcefield and Density-Functional-Theory (DFT) show that carbon atoms remain essentially $sp^2$ coordinated in either bent tubes or tubes pushed by an atomically sharp AFM tip. Subsequent Green's-function-based transport calculations reveal that for *armchair* tubes there is no significant drop in conductance, while for *zigzag* tubes the conductance can drop by several orders of magnitude in AFM-pushed tubes. The effect can be attributed to simple stretching of the tube under tip deformation, which opens up an energy gap at the Fermi surface.

PACS: 61.46.+w, 62.25.+g, 73.63.Fg, 85.35.Kt

---

[*] Corresponding authors, E-mail: 1 amaiti@accelrys.com, 2 svizhenk@nas.nasa.gov

Tremendous potential for technological applications has thrust carbon nanotubes into one of the hottest areas of research activity. This has been fueled by recent experimental breakthroughs in diverse application areas [1], ranging from flat panel displays, to novel microelectronic devices, to hydrogen storage devices, to structural reinforcing agents, to chemical and electromechanical sensors. A pioneering experiment in the last application area involved a metallic nanotube suspended over a 600 nm long trench [2]. When the middle part of such a suspended nanotube was pushed with the tip of an Atomic Force Microscope (AFM), the conductivity was found to decrease by nearly two orders of magnitude for a deformation angle of 15º. The effect was found to be completely reversible, i.e., through repeated cycles of AFM-deformation and tip removal, the electrical conductance displayed a cyclical variation with constant amplitude.

The drop in conductance in the AFM-deformed tube was much higher than the computationally predicted values for tubes *bent* under mechanical duress. Such calculations, using both tight-binding [3] and semi-empirical Extended-Hückel type approaches [4] concluded that even under large bending angles the reduction in electrical conductance was less than an order of magnitude. For AFM-deformed nanotubes, in contrast, O(N) tight-binding calculations [5] show that beyond a critical deformation several C-atoms close to the AFM tip become $sp^3$-coordinated. The $sp^3$ coordination ties up delocalized $\pi$-electrons into localized $\sigma$-states. This would naturally explain the large drop in electrical conductivity, as verified by explicit transport calculations.

Under either bending or pushing by an atomically sharp AFM tip, bond reconstruction, if any, is likely to occur only in the highly deformed, non-straight part of the tube in the middle. This prompted us to use a DFT-based quantum mechanical description of the middle part of the tube (~ 100 atoms), while the long and essentially straight part away from the middle was described accurately using the Universal Forcefield (UFF) [6, 7]. Structures and energetics obtained this way for bent tubes were in good agreement with previous work using an interatomic potential [8]. For the AFM tip-deformed tubes, on the other hand, the situation depended on how the tip was represented. Thus, if the presence of the tip was simulated by constraining a single C-atom on the bottom side of the middle part of the tube, simulations on a (5, 5) armchair led to the development of sp$^3$ coordination between the constrained atom and an atom on the top side at a critical deformation angle of ~ 7º, which destabilized into a complex broken-bond defect at higher deformation angles [9]. On the other hand, a more realistic representation of the AFM-tip by means of an atomically sharp 15-atom Li-needle yielded an sp$^2$-coordinated all-hexagonal tube for deformation angles as high as 25º for the same (5, 5) tube [10]. A large drop in conductance is expected under sp$^3$ coordination and broken bond defects [5, 9]. However, *given the uncertainty of sp$^3$ coordination, can one still expect a significant conductance drop in a tip-deformed, yet sp$^2$-coordinated tube?*

In this Letter, we address the above question by extending the combined DFT-UFF calculations to a (12, 0) metallic zigzag tube. For comparison, we have also considered a (6, 6) armchair tube, which is slightly smaller in diameter, but has the same number of atoms (12) along the tube circumference as the (12, 0) zigzag. The main result of structural relaxation using DFT-UFF is that, sp$^3$ coordination does not happen under either bending or tip-deformation (using an atomically sharp 15-atom Li-tip as in ref. [10]) up to very large angles. Following structural relaxation at each bending and deformation angle, we compute the electronic density of states (DOS), transmission and conductance using the recursive Green's function method [11]. For the armchair tube, the resulting conductance is lowered only by a factor of 1.01 for the tube bent by 40º and a factor of 1.05 for a tube tip-deformed by 25º. Under the same deformations, drop in conductance, for the zigzag tube, is much higher, being a factor of 1.9 under bending, and a remarkable $1.7 \cdot 10^4$ under tip-deformation.

The simulations were carried out on tubes of 2400 atoms, both for the (6, 6) and (12, 0) tubes. Initially the straight tube was relaxed with the UFF. For bending simulations, two halves of the tube were then rotated by equal and opposite angles about an axis perpendicular to the tube and passing through the center of mass of the initial straight tube. At each end of the tube, a contact region defined by a unit cell [12] plus one atomic ring (a total of 36 and 60 atoms for the armchair and the zigzag tube respectively) was then fixed and the whole tube relaxed with the UFF. To simulate AFM-tip-deformation, the 15-atom Li-needle was initially aimed at the center of a hexagon on the bottom-side of the middle part of tube. The Li-needle was then displaced by an amount *d* toward the tube along the needle-axis, resulting in a deformation angle *q* = $tan^{-1}(2d/L)$, *L* being the unstretched length of the tube. The whole tube was then relaxed by UFF keeping the needle atoms and the end contact regions of the tube fixed. Fixing the relative positions of contact region atoms at the same value as in an unstretched tube guarantees that contacts may be approximated by ideal undeformed semi-infinite carbon nanotube leads and that all possible contact modes are coupled to the deformed part of the tube.

Following the UFF relaxation, a cluster of 132 C-atoms for the (6, 6) and a cluster of 144 C-atoms for the (12, 0) were cut out from the middle of the tubes. These clusters, referred to below as the *QM clusters* (plus the 15 Li-tip atoms in tip-deformation simulations) were further relaxed with Accelrys' DFT-code DMol$^3$ [13], with the end atoms of the cluster plus the Li-tip atoms fixed at their respective classical positions. The electronic wave functions in DMol$^3$ were expanded in a double-numeric polarized (DNP) basis set with a real-space cutoff of 4.0 Å. The Hamiltonian was approximated with the Harris functional [14] using a local exchange-correlation potential [15], and the "medium" grid was chosen for numerical integration.

Fig. 1 displays the tip-deformed QM-cluster for the (6, 6) and the (12, 0) tubes at the highest deformation angle of 25º considered in these simulations. Even under such large deformations, there is no indication of $sp^3$ bonding [16], similar to what was observed for the (5, 5) tube in ref. [10]. Although not explicitly shown here, results for *bending* also yield $sp^2$-coordinated all-hexagonal tubes. The absence of $sp^3$ coordination is inferred based on an analysis of nearest neighbor distances of the atoms with the highest displacements, i.e., the ones on the top of the kink in a bent tube, and the ones closest to the Li-tip in a tip-deformed tube. *Although for each of these atoms the three nearest neighbor C-C bonds are stretched to between 1.45-1.75 Å, the distance of the fourth neighbor, required to induce $sp^3$ coordination is greater than 2.2 Å for all tubes in our simulations.* The main difference between a tip-deformed tube versus a bent tube is that there is an overall stretching in the former [17] whereas in the latter case there is no net stretching, and the extra compressive strain on the bottom side is relieved through the formation of a *kink* beyond a critical bending angle.

Following atomic relaxation of the structures, we performed conductance calculations in order to make further predictions on the electromechanical behavior of nanotubes. A coherent conductance was studied within a nearest-neighbor $sp^3$-tight-binding Hamiltonian in a non-orthogonal basis. The parameterization scheme explicitly accounts for effects of strain in the system through a bond-length-dependence of the Hamiltonian and the overlap matrices $H_{ij}$ and $S_{ij}$, as in Ref [18]. We have also checked to confirm that other tight binding parameterizations give qualitatively the same results [19, 20]. First, the retarded Green's function $G^R$ of the whole nanotube was determined by solving the following equation:

$$(E \cdot S_{ij} - H_{ij} - \Sigma_{L,ij} - \Sigma_{R,ij}) G^{R,jk} = \mathbf{d}_i^k, \qquad (1)$$

where $\mathbf{S}_{L,R}$ are the retarded self-energies of the left and the right semi-infinite contacts. The transmission and the electronic density of states (DOS) at each energy were then found [21, 22] from the equations:

$$T(E) = G^{R,ij} \Gamma_{L,jk} G^{A,kl} \Gamma_{R,li}, \qquad (2)$$

$$N_a(E) = -\frac{1}{\pi} \text{Im}\{S_{aj} G^{R,ja}\}, \qquad (3)$$

where $\mathbf{G}_{L,R} = i(\mathbf{S}^R_{L,R} - \mathbf{S}^A_{L,R})$ are the couplings to the left and right leads. Finally, the total conductance of the tube was computed using Landauer-Büttiker formula:

$$G = \frac{2e^2}{h} \int_{-\infty}^{\infty} T(E)(-\frac{\partial f_o}{\partial E}) dE, \qquad (4)$$

where $f_o(E)$ is the Fermi-Dirac function.

Fig. 2 displays the computed conductance (at T = 300K) for the (6, 6) and the (12, 0) tubes as a function of bending and tip-deformation angles. The conductance remains essentially constant for the (6, 6) tube in either bending or tip-deformation simulations. However, for the (12, 0) tube the conductance drops by a factor of 1.9 under bending at $q=40°$, and much more significantly under tip deformation: by ~ 0.3 at 15°, two orders of magnitude at 20°, and 4 orders of magnitude at $q=25°$ [23].

To analyze which part of the zigzag tube is responsible for the conductance drop, we computed the DOS in the vicinity of the Fermi surface. Fig. 3 displays the DOS averaged over 2 unit cells [11] (96 atoms) in three different regions of the AFM-deformed (12, 0) tube for $q=25^o$: (1) undeformed contact, (2) highly deformed tip region, and (3) the uniformly stretched straight regions on either side of the tip-deformed region. DOS in both tip region and the straight part show a bandgap opening, which proves that the conductance drop occurs everywhere in the tube, rather than in the tip-deformed region alone. This has important implications for the application of nanotubes as electromechanical sensors: *given a metallic zigzag nanotube, one could induce a significant conductance drop simply upon uniform stretching*. To check this, we computed the conductance of a uniformly stretched (12, 0) tube as a function of strain, shown in Fig. 4, and compared it to that of AFM-deformed tube from Fig. 2. Both cases show quantitatively the same drastic decay of conductance. The inset in Fig. 4 also shows a band gap opening in transmission in the two cases [24], compared to that of a non-deformed tube.

In order to explain the differences in conductance of the (6, 6) and the (12, 0) tubes as a function of strain, we have analyzed the bandstructures of a metallic zigzag and an armchair nanotube. Starting from the band structure of the 2D graphene [25] under deformation [26], one can derive the following dispersion relations for the crossing subbands within the π-electron approximation (*a'* below is the strained periodic repeat length along the nanotube axis):

$$E(k) = \pm t_2 \{1 + \mathbf{a}^2 - 2\mathbf{a}\cos(\frac{\sqrt{3}ka'}{2})\}^{1/2} \qquad (5a)$$

for a metallic zigzag tube, and

$$E(k) = \pm t_2 \left|1 - 2\mathbf{a}\cos(\frac{ka'}{2})\right| . \qquad (5b)$$

for an armchair tube. Here, $\mathbf{a}=t_1/t_2$ is the ratio of nonequivalent hopping parameters. For a zigzag tube, *a>1*. Consequently, there is no value of *ka'* for which *E(k) =0* (for dispersion (5a)), and a band gap opens up. However, for an armchair tube, *a<1* and one can always find a value of *ka'*, such that *E(k)=0* (for dispersion (5b)), as long as *a>1/2*. The magnitude of the strain-induced bandgap decreases monotonically with increase in chiral angle [26], being maximum for a chiral angle of 0º (zigzag) and gradually reducing to zero at a chiral angle of 30º (armchair). An experiment as in ref. [2] is, thus, expected to show a decrease in conductance as the nanotube is deformed with an AFM tip, for all nanotubes except the armchair tube. The decrease in conductance should be large for the metallic zigzag nanotubes, and small for nanotubes with a chiral angle closer to armchair.

In summary, we find that both under bending and under deformation with an atomically sharp AFM-tip, carbon nanotubes essentially remain all-hexagonal and $sp^2$-coordinated. In the absence of $sp^3$ coordination, armchair tubes remain significantly conducting even at large deformations. However, metallic zigzag tubes display a dramatic drop in conductance, particularly under tip-deformation. A density of states analysis indicates that the conductance drop is distributed over the whole tube, rather than focused in the tip region. This suggests the possibility of designing nanoelectromechanical sensors in which nanotubes are subjected to a uniform tensile strain.

*Acknowledgements*: A.M. would like to acknowledge Accelrys Inc. for its support. A. S. and M. P. A. would like to acknowledge NASA for funding the development of the 2D Quantum Device Simulator used for the conductance calculations in this paper.

Figure captions:

Fig1. DMol$^3$-relaxed Li-tip-deformed QM clusters for: (a) the (6, 6) armchair (132 C-atoms); and (b) the (12, 0) zigzag (144 C-atoms), in side views. The deformation angle is 25º for both tubes. Figs. (c) and (d) are respective views along the tube length, with the Li-tip hidden for clarity.

Fig2. Computed conductance (at T = 300K) for the (6, 6) and (12, 0) nanotubes as a function of: **(top)** bending; and **(bottom)** tip-deformation. Under 40º bending conductance of the zigzag tube drops by a factor of 1.9, while for the armchair tube it drops by only 1.01. Under 25º tip-deformation, conductance of the zigzag tube drops by 4 orders of magnitude, while for the armchair tube it drops only by a factor of 1.05.

Fig. 3. Density of states, averaged over two unit cells (96 atoms) in three different regions of AFM-deformed (12, 0) tube at *q* = 25º. The Fermi surface is at E=0. Nearly equal band gaps open up in the straight part of the tube and in the tip region. Lower DOS in the tip region is due to a larger local straining of C-C bonds close to the AFM-tip.

Fig. 4. Conductance of the uniformly stretched (12, 0) tube compared to that of the tip-deformation case in Fig. 2. Actual angles of tip-deformation are indicated. The % strain for the AFM-deformed tube is computed from the average C-C bond-stretch in the middle of the straight portion of the tube [17]. The inset shows transmission in the vicinity of the Fermi surface (E = 0) for a uniform strain of 10% and a tip-deformation angle of 25º, as compared to the undeformed tube.

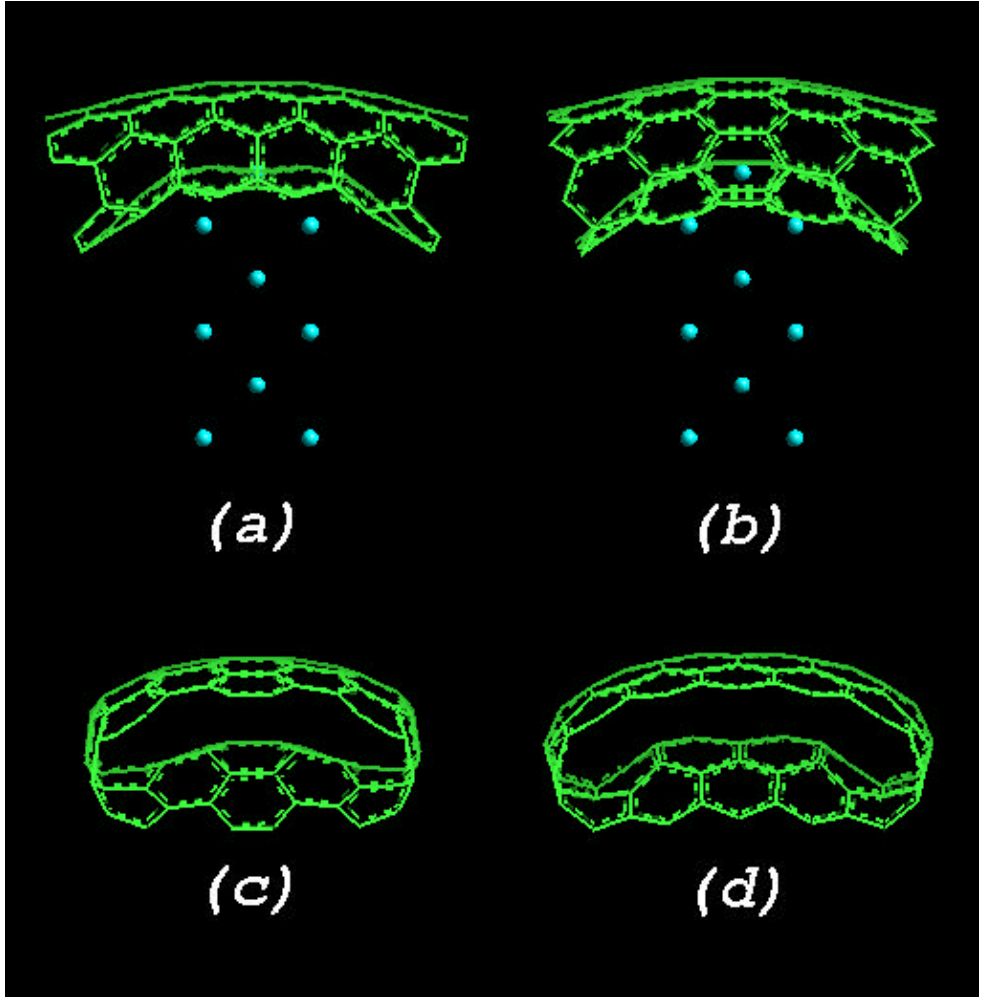

Figure 1

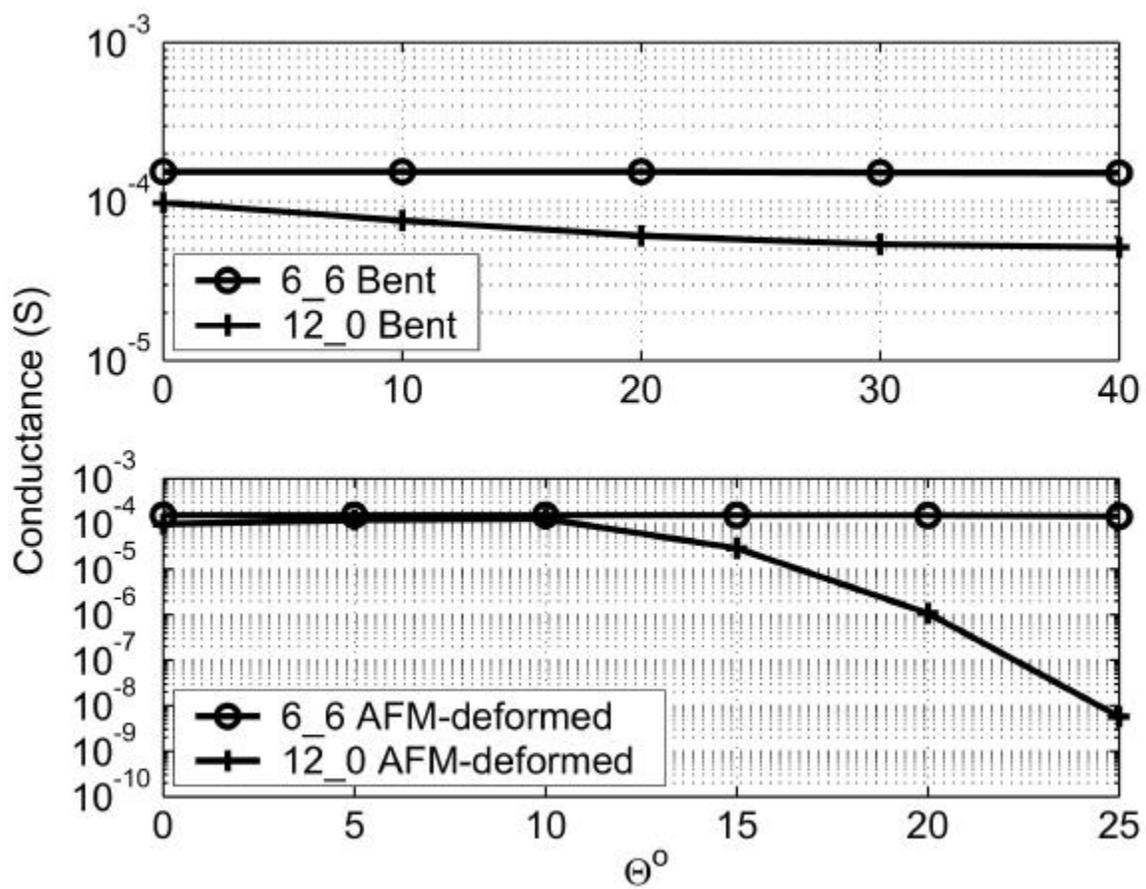

Figure 2

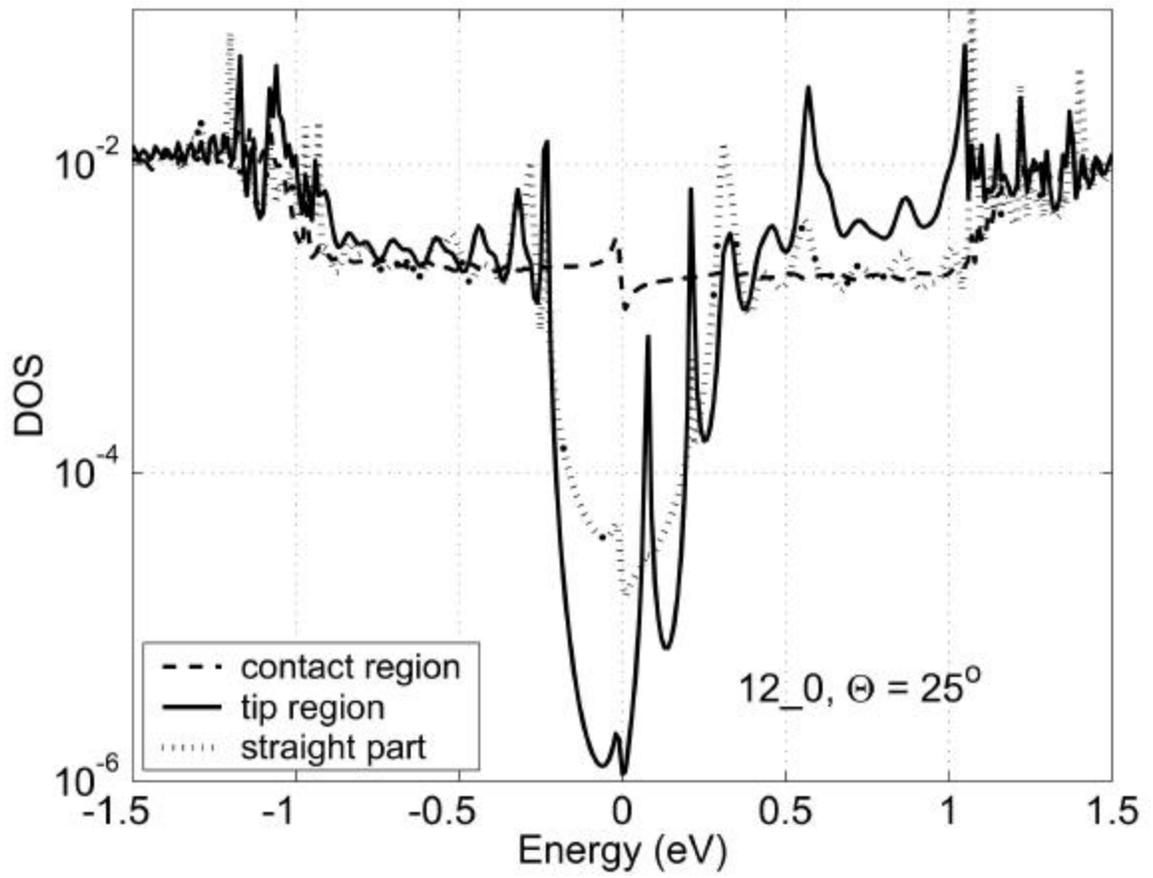

Figure 3

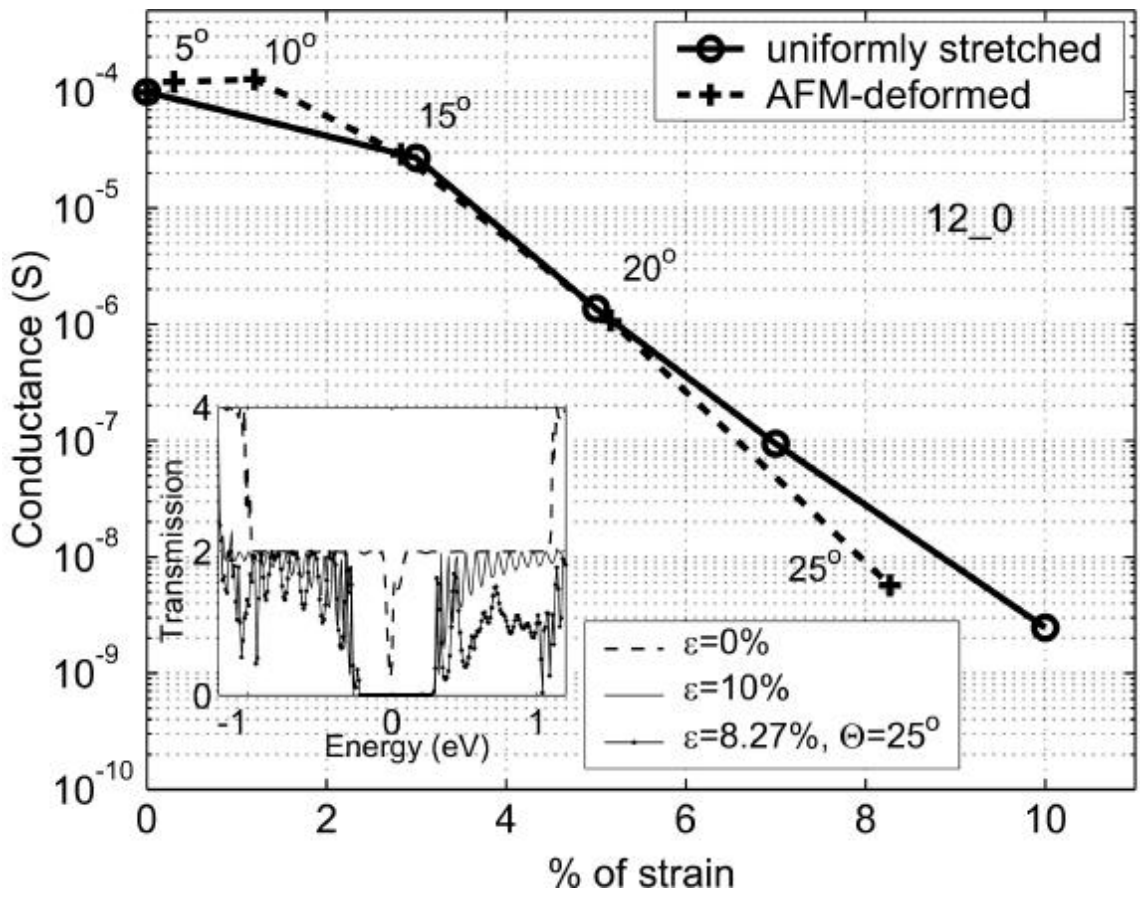

Figure 4